\documentclass[superscriptaddress,reprint,floatfix]{revtex4-1}
\usepackage{graphicx}
\usepackage{amsmath}
\usepackage{amssymb}
\usepackage{bm}
\usepackage[squaren, thinqspace]{SIunits}
\usepackage{xspace}
\usepackage{setspace}
\usepackage{color}

\renewcommand{\degree}{\ensuremath{^\circ}\xspace}

\newcommand{\y}{\ensuremath{\bm{y}}\xspace}
\newcommand{\x}{\ensuremath{\bm{x}}\xspace}

\newcommand{\VANE}{\ensuremath{V_\mathrm{ANE}}\xspace}
\newcommand{\VISHE}{\ensuremath{V_\mathrm{ISHE}}\xspace}

\begin{document}

\title{Local charge and spin currents in magnetothermal landscapes}
\author{Mathias Weiler}
\author{Matthias Althammer}
\author{Franz D. Czeschka}
\author{Hans Huebl}
\author{Martin S. Wagner}
\author{Matthias Opel}
\affiliation{Walther-Mei{\ss}ner-Institut, Bayerische Akademie der Wissenschaften, 85748 Garching, Germany}
\author{Inga-Mareen Imort}
\author{G\"{u}nter Reiss}
\author{Andy Thomas}
\affiliation{Fakult\"{a}t f\"{u}r Physik, Universit\"{a}t Bielefeld, 33615 Bielefeld, Germany}
\author{Rudolf Gross}
\affiliation{Walther-Mei{\ss}ner-Institut, Bayerische Akademie der Wissenschaften, 85748 Garching, Germany}
\affiliation{Physik-Department, Technische Universit\"{a}t M\"{u}nchen, 85748 Garching, Germany}
\author{Sebastian T. B. Goennenwein}
\email[Electronic address:]{goennenwein@wmi.badw.de}
\affiliation{Walther-Mei{\ss}ner-Institut, Bayerische Akademie der Wissenschaften, 85748 Garching, Germany}


\begin{abstract}
A scannable laser beam is used to generate local thermal gradients in metallic (Co$_2$FeAl) or insulating (Y$_3$Fe$_5$O$_{12}$) ferromagnetic thin films. We study the resulting local charge and spin currents that arise due to the anomalous Nernst effect (ANE) and the spin Seebeck effect (SSE), respectively. In the local ANE experiments, we detect the voltage in the Co$_2$FeAl thin film plane as a function of the laser spot position and external magnetic field magnitude and orientation. The local SSE effect is detected in a similar fashion by exploiting the inverse spin Hall effect in a Pt layer deposited on top of the Y$_3$Fe$_5$O$_{12}$. Our findings establish local thermal spin and charge current generation as well as spin caloritronic domain imaging.
\end{abstract}

\maketitle


Spin caloritronic effects have been extensively studied using integral (homogeneous) thermal gradients~\cite{Nernst:1887,Harmann:1967}. In ferromagnetic conductors exposed to a thermal gradient in the Nernst geometry, one observes the anomalous Nernst effect (ANE), which describes the occurrence of an electric field $\bm{E}_\mathrm{ANE}\propto -\bm{M} \times \nabla T$, perpendicular to both, the temperature gradient $\nabla T$ and the magnetization $\bm{M}$. The anomalous Nernst effect has been studied in a variety of ferromagnetic thin film metals~\cite{Vasil'eva:1972, Miyasato:2007}, oxides~\cite{Suryanarayanan:1999, Miyasato:2007}, spinels~\cite{Lee:2004, Miyasato:2007,Hanasaki:2008} and diluted magnetic semiconductors~\cite{Pu:2008}. In analogy to charge-based caloritronic effects, the recently discovered spin Seebeck effect (SSE)~\cite{Uchida:2008} describes the generation of a spin current $\bm{J}_\mathrm{s}$ parallel to an applied temperature gradient $\nabla T$ in ferromagnetic materials. $\bm{J}_\mathrm{s}$ can be detected all electrically by exploiting the inverse spin Hall effect~\cite{Hirsch:1999, Saitoh:2006} in a normal metal (N) deposited on top of the ferromagnet (FM). In the longitudinal spin Seebeck configuration~\cite{Uchida2:2010}, $\nabla T$ is applied along the FM/N hybrid normal, resulting in an electric field $\bm{E}_\mathrm{ISHE}\propto \bm{J}_\mathrm{s} \times \bm{\sigma}$. Here, $\bm{\sigma}\parallel \bm{M}$ is the spin polarization, such that the symmetry of the SSE is identical to the ANE with respect to $\bm{M}$ and $\nabla T$. Spin Seebeck measurements have been carried out in ferromagnetic metals~\cite{Uchida:2008, Bosu:2011}, diluted magnetic semiconductors~\cite{Jaworski:2010} and magnetic insulators~\cite{Uchida:2010}. The interplay of spins and temperature leads to further intriguing effects such as the spin Peltier effect~\cite{Gravier:2005, Gravier:2006}, thermal spin torque~\cite{Slonczweski:2010, Yu:2010}, or thermally driven spin injection~\cite{Slachter:2010, leBreton:2011}. However, in all spin caloritronic experiments mentioned above, homogeneous temperature gradients were applied. In order to establish the interplay between temperature gradients and spin degrees of freedom also on the length scale of the magnetic microstructure, temperature gradients changing on such length scales are mandatory. Here, we therefore use a focussed, scanning laser beam to generate a local temperature gradient perpendicular to a thin film sample plane, and perform a spatially resolved study of the resulting spin caloritronic effects. Our findings demonstrate that spatially confined thermal gradients allow for the generation of local, bipolar and magnetically controllable electric fields or spin currents that can be used to, e.g., electrically image the magnetic microstructure in ferromagnetic metals and insulators.

%
\begin{figure}
  \includegraphics[]{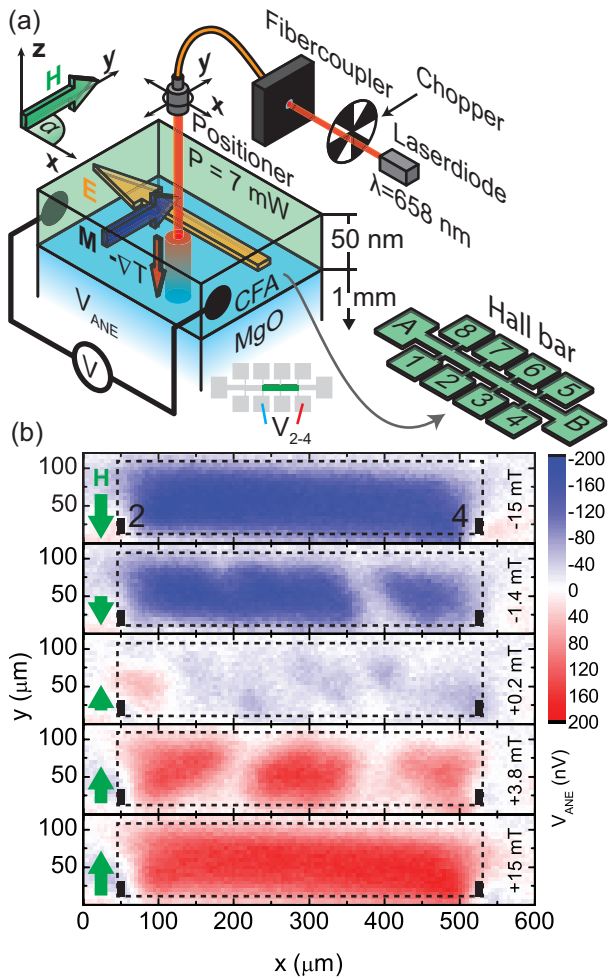}\\
  \caption{(a) The scannable laser beam generates a local temperature gradient $\nabla T$ normal to the ferromagnetic thin film plane. The dc voltage \VANE which arises due to the anomalous Nernst effect depends on the local magnetization $\bm{M}$ at the position ($x,y$) of the laser beam. All investigated samples are patterned into \unit{80}{\micro\meter} wide and \unit{900}{\micro\meter} long Hall bars with contacts labeled as sketched. (b) \VANE determined between contacts 2 and 4 as a function of the laser-spot position ($x,y$) and the external magnetic field magnitude $\mu_0 H$ in a \unit{50}{\nano\meter} thick Co$_2$FeAl (CFA) film.}\label{fig:spatial}
\end{figure}
In these spatially resolved experiments, the local anomalous Nernst effect in a conductive ferromagnetic thin film results in an electric field
\begin{equation}\label{eq:ANE}
    \bm{E}_\mathrm{ANE}(x,y)=- N \mu_0 \bm{M}(x,y) \times \nabla T(x,y)\;,
\end{equation}
at position $(x,y)$ with the Nernst coefficient $N$. In samples consisting of a ferromagnetic insulator/normal metal bilayer exposed to a local temperature gradient, the spin Seebeck and inverse spin Hall effect yield a local electric field
\begin{equation}\label{eq:SSE}
    \bm{E}_\mathrm{ISHE}(x,y)=- S \bm{\sigma}(x,y) \times \nabla T(x,y)\;,
\end{equation}
defined analogous to the integral expression found in Ref.~\cite{Jaworski:2010} with the phenomenological spin Seebeck coefficient $S$ and the spin polarization vector $\bm{\sigma}=\bm{M}/M_\mathrm{s}$, with the saturation magnetization $M_\mathrm{s}$. Comparing Eq.~\eqref{eq:ANE} and Eq.~\eqref{eq:SSE}, it is evident that $\bm{E}_\mathrm{ANE}$ and $\bm{E}_\mathrm{ISHE}$ bear identical symmetry. Hence both, $\bm{E}_\mathrm{ANE}(x,y)$ and $\bm{E}_\mathrm{ISHE}(x,y)$ can be detected in an identical fashion, enabling a spatially resolved investigation of charge and spin currents in a magnetothermal experiment. It is important to note that $\bm{E}_\mathrm{ANE}(x,y)$ and $\bm{E}_\mathrm{ISHE}(x,y)$ are local electric fields, determined by the magnetic properties and temperature gradient at position $(x,y)$. In conductive ferromagnets with $N\neq0$, a spatially confined $\nabla T(x,y)$ will thus evoke a local $ \bm{E}_\mathrm{ANE}(x,y)$. Its magnitude and polarity are controllable in situ by manipulating $\bm{M}(x,y)$. Vice versa, $\bm{E}_\mathrm{ANE}(x,y)$ can be used to electrically read out the magnetization $\bm{M}(x,y)$ with full 360\degree confidence, i.e., to electrically image the magnetic microstructure by scanning $\nabla T(x,y)$ across the sample. Identical considerations apply in magnetic insulator / normal metal bilayers where $\bm{E}_\mathrm{ISHE}(x,y)$ is generated in the presence of a temperature gradient. $\bm{E}_\mathrm{ISHE}(x,y)$ is, however, caused by a local spin current $\bm{J}_\mathrm{s}(x,y)$. Hence the detection of $\bm{E}_{\mathrm{ISHE}}(x,y)$ not only allows to electrically detect magnetic texture in a ferromagnetic insulator, but even enables a spatial mapping of spin currents.

We first demonstrate magnetothermal domain imaging in a conductive ferromagnetic thin film via the anomalous Nernst effect. To this end, the setup depicted schematically in Fig.~\ref{fig:spatial}(a) is used (for details, see Appendix A). It is operated at room temperature for all measurements discussed in this work. The light beam emitted by a laser diode is coupled into an optical fiber and focussed onto the sample surface at position $(x,y)$ by means of a scannable collimator. Since the sample at least partially absorbs the laser light, its intensity and thus the energy deposited decrease as a function of depth. Hence, the energy absorption profile of the laser beam into the film thickness gives rise to a thermal gradient $\nabla_\mathrm{z} T(x,y)$ perpendicular to the sample plane, laterally confined to a region around the position $(x,y)$ of the laser spot~\cite{Reichling:1994}. This thermal gradient gives rise to a local electric field $\bm{E}_\mathrm{ANE}(x,y)$ (cf. Eq.~\eqref{eq:ANE}). Temperature gradients within the sample plane are radially symmetric and their contributions to magnetothermal effects thus cancel out. We use a ferromagnetic Co$_2$FeAl thin film deposited on a MgO substrate (see Appendix A). The film is patterned into the Hall bar geometry shown in Fig.~\ref{fig:spatial}(a). The magnetic microstructure of this particular sample is known from magneto-optical Kerr effect measurements~\cite{Weiler:2011}.

Figure~\ref{fig:spatial}(b) shows the dc voltage \VANE recorded between the contacts 2 and 4 which are separated by approximately \unit{460}{\micro\meter}. For each value of the in-plane magnetic field $\mu_0H$ applied at an angle $\alpha=90\degree$ to the $\bm{x}$ axis, we scanned the laser beam over the central Hall bar area and recorded $\VANE(x,y)$ as a function of the laser spot position ($x,y$). The small full rectangles indicate the location of the used electric contacts and the dashed rectangle depicts the region on the main Hall bar enclosed by said contacts. At $\mu_0H=\unit{-15}{\milli\tesla}$ ($\bm{H}\parallel-\bm{y}$ as indicated by the solid arrow to the left) we observe a voltage $\VANE\approx\unit{-150}{\nano\volt}$ in the Hall bar region independent of the laser spot position ($x,y$). We attribute $\VANE\propto E_\mathrm{x}$ to the anomalous Nernst effect defined in Eq.~\eqref{eq:ANE}. At $\mu_0H=\unit{-15}{\milli\tesla}$ the film is in magnetic saturation with $\bm{M} \parallel \bm{H}$ as shown later. Hence, no magnetic microstructure is present and $\VANE(x,y)$ does not change as a function of $x$ and $y$. As the magnetic field magnitude is decreased to $\mu_0H=\unit{-1.4}{\milli\tesla}$, magnetic domain formation is evident from the \VANE map and at $\mu_0H=\unit{+0.2}{\milli\tesla}$ \VANE vanishes in the major part of the Hall bar, indicating that $\bm{M}$ is oriented (anti-)parallel to $\bm{x}$, such that $\left(\bm{M} \times \nabla T\right)\cdot\bm{x}=0$. Note that Co$_2$FeAl has cubic magnetic anisotropy. As a consequence, the magnetic reversal proceeds via two 90\degree switches~\cite{Weiler:2011}. Upon increasing the external magnetic field to $\mu_0H=\unit{+3.8}{\milli\tesla}$, domains exhibiting $\VANE>0$ become visible. In magnetic saturation at $\mu_0H=\unit{+15}{\milli\tesla}$, $\VANE\approx\unit{+150}{\nano\volt}$ in the entire Hall bar region. The sign reversal of \VANE with the reversal of the direction of $\bm{H}$ (and thus $\bm{M}$) is a clear indication that the observed \VANE indeed is caused by a term $\nabla T \times \bm{M}$, which allows to rule out all field-symmetric thermopower effects as the cause of the observed voltage. This is completely analogous to the Nernst signal in the mixed state of superconductors which changes sign upon switching the direction of the flux lines~\cite{Ri:1994}. Furthermore, \VANE is a local voltage as can be seen by the fact that $\VANE(x<\unit{50}{\micro\meter},y)=0$ and $\VANE(x>\unit{550}{\micro\meter},y)=0$. For these $x$, the laser still impinges on the main Hall bar, it is however on either side of both contacts, rendering them at identical electrical potential (see Appendix C).

%
\begin{figure}
  \includegraphics[]{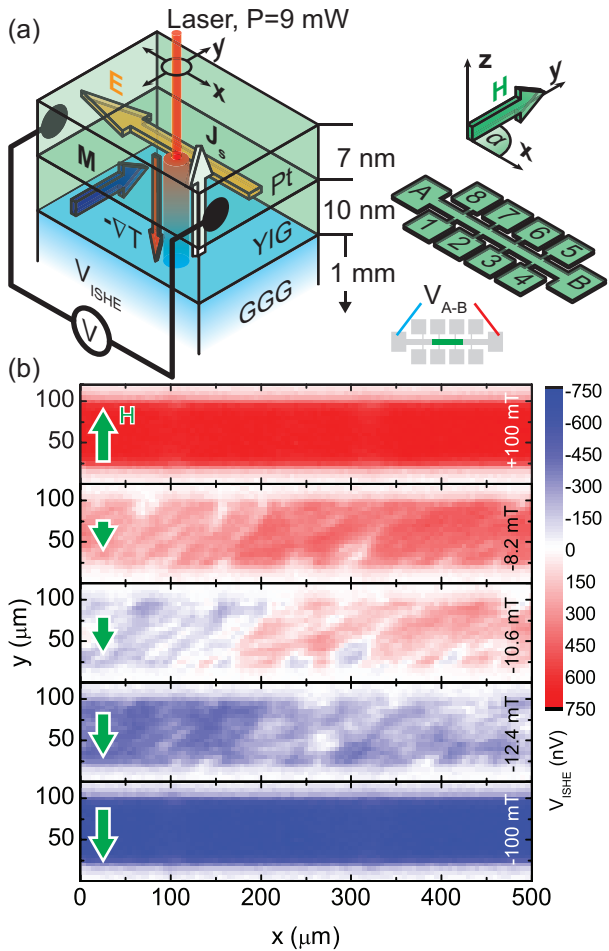}\\
  \caption{(a) Sample schematics of GGG/YIG/Pt sample. The spin Seebeck effect yields a pure, local spin current $\bm{J}_\mathrm{s}$ along $\nabla T$ in YIG. $\bm{J}_\mathrm{s}$ depends on the local magnetization $\bm{M}(x,y)$ and is detected using the inverse spin Hall effect in Pt, which gives rise to a dc voltage \VISHE. (b) \VISHE determined between contacts A and B as a function of the laser-spot position $(x,y)$ and the external magnetic field magnitude $\mu_0 H$ applied along \y in a \unit{10}{\nano\meter} YIG / \unit{7}{\nano\meter} Pt hybrid sample.}\label{fig:SSE}
\end{figure}

We now turn to the generation and detection of local spin currents via the longitudinal spin Seebeck effect in a ferromagnetic insulator exposed to magnetothermal landscapes. We employ a \unit{10}{\nano\meter} thick Y$_3$Fe$_5$O$_{12}$ (YIG) film grown onto on a Gd$_3$Ga$_5$O$_12$ (GGG) substrate (see Appendix A). The YIG thin film was covered in situ by a \unit{7}{\nano\meter} thick Pt film to take advantage of the inverse spin Hall effect for an all electrical detection of local spin currents. The YIG/Pt hybrid was patterned into the same Hall bar geometry as the Co$_2$FeAl sample. A schematic view of the sample and setup (which is identical to that used for the ANE measurements) is shown in Fig.~\ref{fig:SSE}(a). Upon application of a temperature gradient along the hybrid normal, the longitudinal SSE~\cite{Uchida2:2010} yields a pure spin current $\bm{J}_\mathrm{s}$ in the YIG film parallel to $\nabla T$ which can be detected by exploiting the inverse spin Hall effect in the Pt layer. Note that YIG is an electrical insulator, such that it does not show an anomalous Nernst effect. The Pt layer serves not only as a spin current detector but furthermore as an optical absorber of the laser light.

In Fig.~\ref{fig:SSE}(b) we present a spatially resolved measurement of $\VISHE=V_\mathrm{A-B}$ as a function of $H$. Figure~\ref{fig:SSE}(b) thus represents a map of magnetic domains in the ferromagnetic insulator YIG, detected by local electric fields in a Pt layer deposited on top. At $\mu_0 H=\unit{\pm100}{\milli\tesla}$ (top and bottom panel) the YIG thin film is in a single domain state with $\bm{M} \parallel \bm{H}$. As $\bm{H}$ is applied along \y, we can observe $\bm{E}_\mathrm{ISHE}$ along \x by probing \VISHE (cf. Eq.~\eqref{eq:SSE}). As we used contacts A and B for recording \VISHE, the laser spot position is located between the used contacts for all values of $x$ in Fig.~\ref{fig:SSE}(b). For all values of $y$ where the laser impinges on the \unit{80}{\micro\meter} wide Hall bar, $\VISHE=\unit{+640}{\nano\volt}$ in magnetic saturation with $\mu_0 H=\unit{+100}{\milli\tesla}$. Magnetic texture can be observed during the magnetic field sweep in the images recorded with $\mu_0 H=\unit{+8.2}{\milli\tesla}$ to $\mu_0 H=\unit{-12.4}{\milli\tesla}$ (middle panels). As our YIG (111) films show only very small magnetic anisotropy in the film plane the magnetic domain pattern is more complex than that observed in Co$_2$FeAl. We note that - while our YIG thin films are electrically insulating - magnetotransport measurements on our YIG/Pt samples showed an anisotropic magnetoresistance $\Delta R/R \approx 7\times10^{-4}$ (see Appendix D) attributed to induced magnetic moments in the Pt thin film close to the interface~\cite{Wilhelm:2000}. An interpretation of \VISHE on the basis of Nernst effects in Pt would require implausible Nernst coefficients (see Appendix D). We thus attribute the magnetothermal voltage observed to the longitudinal SSE.

%
\begin{figure}
  \includegraphics[]{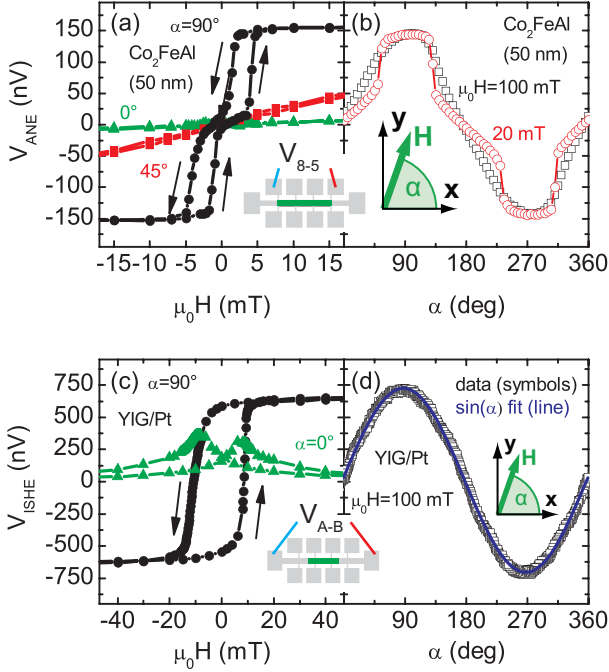}\\
  \caption{(a) Averaged \VANE in Co$_2$FeAl using the region of interest and contacts shown in the inset as a function of $\mu_0H$ for $\alpha\in\left\{0\degree, 45\degree, 90\degree\right\}$ (triangles, squares, circles). (b) Sine like angular dependency of \VANE on $\alpha$ at $\mu_0H=\unit{100}{\milli\tesla}$ (squares). At $\mu_0H=\unit{20}{\milli\tesla}$ (circles), the abrupt $\bm{M}$ switching across the magnetically hard axes (45\degree and 135\degree) becomes visible as steps in \VANE. (c)  Averaged \VISHE in YIG/Pt within the main Hall bar region as a function of $\mu_0H$ for $\alpha\in\left\{0\degree, 90\degree\right\}$ (triangles, circles). (d) \VISHE as a function of $\alpha$ (symbols). The solid curve is a fit to $\sin(\alpha)$ showing excellent agreement with data.}\label{fig:integral}
\end{figure}

To quantitatively compare our spatially resolved magnetothermal voltages with the known integral properties, we average \VANE and \VISHE within the illuminated regions. Such integral measurements are presented in Fig.~\ref{fig:integral}, where the insets depict the regions of interest (ROI) used for averaging. In Fig.~\ref{fig:integral}(a), \VANE in Co$_2$FeAl is shown as a function of the external magnetic field magnitude $\mu_0H$ for three different orientations $\alpha$ of the external magnetic field in the plane of the film. For $\alpha=90\degree$ (circles), we observe a double step switching behavior indicating cubic magnetic anisotropy. For large $H$ applied at $\alpha=90\degree$, $\bm{M}$ is oriented perpendicular to the main Hall bar. Hence, the generated electric field $\bm{E}$ is parallel to the main Hall bar and can be probed by the potential difference between the used contacts. For $\alpha=45\degree$, \VANE is smaller, as only the projection of $\bm{E}$ on the $\bm{x}$ direction is probed. Furthermore, at this value of $\alpha$, $\bm{H}$ is oriented along a magnetically hard axis of the Co$_2$FeAl film, so hysteresis is minimal. Finally, for $\alpha=0\degree$, \VANE vanishes because $\bm{E}$ is generated along the $\bm{y}$-direction and thus cannot be detected by voltage contacts aligned along the $\bm{x}$ direction. The evolution of \VANE as a function of $\bm{H}$ orientation is shown in more detail in Fig.~\ref{fig:integral}(b), where \VANE data recorded during a rotation of $\mu_0H=\unit{100}{\milli\tesla}$ within the film plane are depicted (squares). A dependence $\VANE \propto \sin(\alpha)$ is observed in agreement with the cross product found in Eq.~\eqref{eq:ANE}. In an analogous experiment with $\mu_0H=\unit{20}{\milli\tesla}$, a similar behavior is found (circles). However, at the four magnetically hard axes along 45\degree, 135\degree, 225\degree and 315\degree the magnetization switches abruptly, as evident from the steps in \VANE at these orientations. This shows that the anomalous Nernst effect measurements can be used to probe magnetic anisotropy in the same fashion as in angular dependent magneto resistance (ADMR) measurements~\cite{Limmer:2006}, but with spatial resolution.

Figure~\ref{fig:integral}(c) shows $V_\mathrm{A-B}=\VISHE \propto E_\mathrm{ISHE}$ obtained in the YIG/Pt bilayer as a function of the external magnetic field for $\alpha=90\degree$ (circles) and $\alpha=0\degree$ (triangles). As expected from Eq.~\eqref{eq:SSE}, and following the same line of arguments as for \VANE, we observe an antisymmetric \VISHE vs. $H$ behavior for $\alpha=90\degree$ while for $\alpha=0\degree$, \VISHE vanishes for large values of $H$. In Fig.~\ref{fig:integral}(d) we present \VISHE data as a function of $\alpha$ (symbols) recorded with $\mu_0H=\unit{100}{\milli\tesla}$ together with a fit to $\sin(\alpha)$ (line). The excellent agreement between fit and data corroborates the cross product in Eq.~\eqref{eq:SSE}. Thus, by exploiting the SSE it is possible to perform spatially resolved ADMR-like measurements in magnetic insulators.

Upon calculating the temperature gradient $\nabla T$ evoked by the laser heating, we can quantify the anomalous Nernst coefficient $N$ and the spin Seebeck coefficient $S$ of the investigated samples. For Co$_2$FeAl, such a quantitative evaluation is not straightforward, since neither the optical properties nor the Nernst coefficient have been reported. We thus performed further \VANE measurements in a Ni thin film sample (see Appendix C). For Ni, we calculated $\nabla T=\unit{-1.4}{\kelvin\per\micro\meter}$ (see Appendix B). Using a saturation magnetization $M_\mathrm{s}=\unit{370}{\kilo\ampere\per\meter}$ obtained by SQUID magnetometry~\cite{Weiler:2009} and the experimentally measured $E=\unit{87}{\milli\volt\per\meter}$ (see Appendix C), we obtain the Nernst coefficient $N_\mathrm{Ni}\approx\unit{1.3\times10^{-7}}{\volt\per\kelvin\tesla}$ which is lower than values $N_\mathrm{Ni}\approx\unit{5\times10^{-7}}{\volt\per\kelvin\tesla}$ found for bulk Ni~\cite{Smith:1911} at room temperature. A comparable reduction of a caloritronic property with respect to bulk material was recently reported for the Seebeck coefficient in Ni thin films~\cite{Avery:2011}. Assuming comparable optical and thermal properties for Co$_2$FeAl and thus $\nabla T=\unit{-1.4}{\kelvin\per\micro\meter}$ in a \unit{80}{\nano\meter} thick Co$_2$FeAl sample with $M_\mathrm{s}=\unit{1050}{\kilo\ampere\per\meter}$ obtained by SQUID magnetometry, we estimate $N_\mathrm{CFA}\approx\unit{9.5\times10^{-8}}{\volt\per\kelvin\tesla}$. We now turn to the longitudinal spin Seebeck coefficient of the YIG/Pt bilayer. Using a calculated mean temperature gradient of $\nabla T=\unit{-8.7}{\kelvin\per\micro\meter}$ (see Appendix B) in the YIG thin film, we obtain a spin Seebeck coefficient $S=\unit{5.9\times10^{-8}}{\volt\per\kelvin}$, compared to $S=\unit{1\times10^{-7}}{\volt\per\kelvin}$ found in~\cite{Uchida2:2010}. We assume that the small difference in $S$ is due to different interfaces of Pt and YIG as well as different Pt thicknesses. Taken together, the above results show that, by exploiting the longitudinal SSE in a magnetothermal landscape, a determination of the magnetic microstructure by means of an integral voltage measurement is possible even in a ferromagnetic insulator. Furthermore, the spatially resolved SSE and ANE open an avenue for the local generation of pure spin or bipolar charge currents with magnetically selectable (spin) polarization.

In conclusion, our results show that spatially resolved spin caloritronics are a viable path for the use of heat landscapes in spintronic applications. In conductive ferromagnetic thin films, we demonstrated that a spatially confined thermal gradient allows for the generation of a local, magnetically controllable, electric field via the anomalous Nernst effect. Our results furthermore suggest that in magnetic insulator/normal metal hybrids, a spatially confined temperature gradient gives rise to local, pure spin currents with magnetically selectable spin polarization due to the longitudinal spin Seebeck effect. This opens exciting perspectives for the generation and use of pure spin currents, both in basic research and in applications.

Financial support from the DFG via SPP 1538 "Spin Caloric Transport", Project no GO 944/4-1 and the German Excellence Initiative via the "Nanosystems Initiative Munich" (NIM) is gratefully acknowledged.

\appendix
\section{Experimental setup and sample preparation}
A laser diode with $\lambda=\unit{658}{\nano\meter}$ and $P_0=\unit{40}{\milli\watt}$ was coupled into a single-mode optical fiber which terminated in a fiber collimator and was focused with a $f=\unit{11}{\milli\meter}$ lens onto the sample surface. The laser spot size was diffraction limited at $\unit{10}{\micro\meter}$ and all laser powers mentioned in the main text were measured directly at the sample position. The fiber collimator was mounted in a motorized xyz flexure stage, allowing for 3D-positioning in a travel range of \unit{4}{\milli\meter} with a repeatable accuracy of \unit{500}{\nano\meter}. An optical chopper operated at $\nu=\unit{817 - 853}{\hertz}$ and lock-in detection using the differential input of a Stanford Research SR830 instrument was used to record \VANE and \VISHE. The magnetic field $\mu_0H\leq\unit{100}{\milli\tesla}$ was provided by a 2D vector magnet at any orientation $\alpha$ within the sample plane. The Co$_2$FeAl thin films were prepared as described in~\cite{Weiler:2011}, the Y$_3$Fe$_5$O$_{12}$ (111) thin films were prepared using pulsed laser deposition on Gd$_3$Ga$_5$O$_12$ substrates, the Pt was deposited using electron beam evaporation. The Hall bar was defined in a photolithography / etching process. The polycrystalline Ni film was prepared by photolithography and electron beam evaporation on a \unit{1}{\milli\meter} thick MgO substrate followed by a lift-off process to define the Hall bar geometry.

\section{Thermal landscape generation}
\begin{figure}
  \includegraphics[]{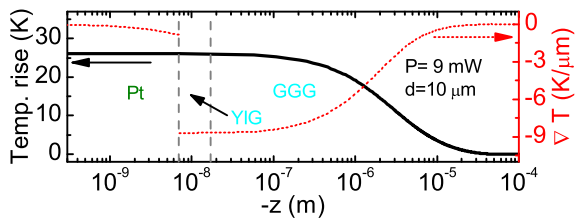}\\
  \caption{Calculated temperature profile (solid line, left scale) and gradient (dotted line, right scale) in a Pt (\unit{7}{\nano\meter}) on YIG  (\unit{10}{\nano\meter}) / GGG (\unit{1}{\milli\meter}) sample in the direction of the film normal $\bm{z}$ for an incident laser power $P=\unit{9}{\milli\watt}$ and a gaussian laser profile with a diameter of \unit{10}{\micro\meter}. The gradient in the YIG film at the YIG/Pt interface along $\bm{z}$ amounts to \unit{-8.7}{\kelvin\per\micro\meter}.}\label{fig:gradTSSE}
\end{figure}
In order to quantitatively determine either the Nernst coefficient $N$ or the longitudinal spin Seebeck coefficient $S$ using the setup described in the main text, a quantitative knowledge of the temperature gradient driving the magnetothermal voltages \VANE and \VISHE is essential. The temperature gradient induced by a laser beam with Gaussian profile impinging on a multilayered sample can be calculated analytically as shown by Reichling and Gr\"{o}nbeck~\cite{Reichling:1994}. Since not all required material parameters are known for Co$_2$FeAl, we carried out the calculation for MgO/Ni and GGG/YIG/Pt.

For the determination of $S$ we calculated the temperature gradient in the GGG/YIG/Pt sample following the approach in~\cite{Reichling:1994}. The result of the calculation for a laser chopping frequency of $\unit{853}{\hertz}$ and an incident laser power $P=\unit{9}{\milli\watt}$ is shown in Fig.~\ref{fig:gradTSSE}. We plot $T$ as a function of the position $z$ along the film normal at the center of the laser beam (left scale), with $z<0$ corresponding to positions within the sample. The right scale shows the temperature gradient $\nabla T(z)$. We assumed identical thermal and optical properties of YIG and GGG and used the following material parameters: $a_\mathrm{Pt}=\unit{82\times10^6}{\per\meter}$ and $a_\mathrm{YIG}=0$ for the absorption coefficients, $R_\mathrm{Pt}=0.68$ and $R_\mathrm{YIG}=0.08587$ for the reflectivities. For the thermal properties we used $\kappa_\mathrm{Air}=\unit{0.02}{\watt\per\kelvin\meter}$, $\kappa_\mathrm{Pt}=\unit{71.6}{\watt\per\kelvin\meter}$ and $\kappa_\mathrm{YIG}=\unit{7}{\watt\per\kelvin\meter}$~\cite{Slack:1971} for the thermal conductivities and $c_\mathrm{Air}=\unit{1000}{\joule\per\kilogram\kelvin}$, $c_\mathrm{Pt}=\unit{130}{\joule\per\kilogram\kelvin}$ and $c_\mathrm{YIG}=\unit{570}{\joule\per\kilogram\kelvin}$ for the heat capacities. Finally, $\rho_\mathrm{Air}=\unit{1.2}{\kilogram\per\meter\cubed}$, $\rho_\mathrm{Pt}=\unit{21450}{\kilogram\per\meter\cubed}$ and $\rho_\mathrm{YIG}=\unit{5245}{\kilogram\per\meter\cubed}$ were taken as the respective densities. We obtain a temperature gradient of approximately $\nabla T=\unit{-8.7}{\kelvin\per\micro\meter}$ in the YIG close to the Pt interface. In the Pt layer, the mean temperature gradient amounts to $\nabla T=\unit{-0.6}{\kelvin\per\micro\meter}$.

\begin{figure}[t]
  \includegraphics[]{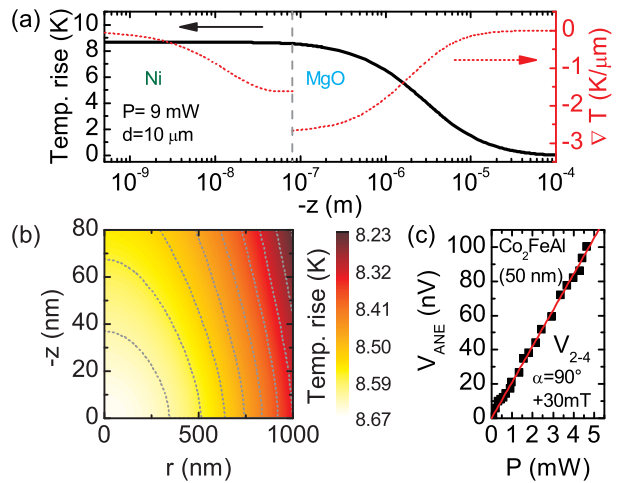}\\
  \caption{(a) Calculated temperature profile (solid line, left scale) and temperature gradient (dotted line, right scale) in a Ni (\unit{80}{\nano\meter}) on MgO (\unit{1}{\milli\meter}) sample in the direction of the film normal $\bm{z}$ for an incident laser power $P=\unit{9}{\milli\watt}$ and a gaussian laser profile with a diameter of \unit{10}{\micro\meter}. The mean gradient in the Ni film along $\bm{z}$ amounts to \unit{-1.4}{\kelvin\per\micro\meter}. (b) The radially symmetric temperature profile in the Ni film as a function of depth $z$ and radial coordinate $r$. (c) \VANE scales linearly with laser power (symbols are measured data).}\label{fig:gradANE}
\end{figure}

For the determination of $N$ we calculated the temperature profile considering a \unit{80}{\nano\meter} thick Ni film on a \unit{1}{\milli\meter} thick MgO substrate. We used the following material parameters: $a_\mathrm{Ni}=\unit{73\times10^6}{\per\meter}$ and $a_\mathrm{MgO}=0$ for the absorption coefficients, $R_\mathrm{Ni}=0.66071$ and $R_\mathrm{MgO}=0.071970$ for the reflectivities, with values taken from Ref.~\cite{Palik:1985} at $\lambda=\unit{658}{\nano\meter}$. For the thermal properties we used $\kappa_\mathrm{Air}=\unit{0.02}{\watt\per\kelvin\meter}$~\cite{Vargaftik:1993}, $\kappa_\mathrm{Ni}=\unit{90}{\watt\per\kelvin\meter}$~\cite{Farrell:1969} and $\kappa_\mathrm{MgO}=\unit{53}{\watt\per\kelvin\meter}$~\cite{Itatani:2006} for the thermal conductivities and $c_\mathrm{Air}=\unit{1000}{\joule\per\kilogram\kelvin}$~\cite{Slocum:1956}, $c_\mathrm{Ni}=\unit{439}{\joule\per\kilogram\kelvin}$~\cite{Lide:2003} and $c_\mathrm{MgO}=\unit{960}{\joule\per\kilogram\kelvin}$~\cite{Watanabe:1993} for the heat capacities. Finally, $\rho_\mathrm{Air}=\unit{1.2}{\kilogram\per\meter\cubed}$, $\rho_\mathrm{Ni}=\unit{8900}{\kilogram\per\meter\cubed}$ and $\rho_\mathrm{MgO}=\unit{3580}{\kilogram\per\meter\cubed}$ were taken as the respective densities.
The calculated temperature $T$, considering the same laser beam profile as above with $d=\unit{10}{\micro\meter}$ diameter, a laser power of $P=\unit{9}{\milli\watt}$ and a chopping frequency of \unit{817}{\hertz} are shown in Fig.~\ref{fig:gradANE}(a). The temperature gradient $\nabla T$ along $\bm{z}$ (right scale) is in average \unit{-1.4}{\kelvin\per\micro\meter} in the Ni thin film. Figure~\ref{fig:gradANE}(b) shows the temperature in the Ni film as a function of $z$ and the radial coordinate $r$, with $r=0$ corresponding to the center of the laser beam. The temperature gradient along $z$ is roughly one order of magnitude higher than that along $r$. Furthermore, as the lateral temperature gradient is radially symmetric, its contributions to the conventional Seebeck effect cancel out and thus need not to be considered in the interpretation of data presented here.

Both, the ANE and the SSE scale linearly with the temperature gradient (cf. Eq.~\eqref{eq:ANE} and~\eqref{eq:SSE}). We experimentally checked the dependence of the effects on the incident laser power which again is proportional to $\nabla T$~\cite{Reichling:1994}. Fig.~\ref{fig:gradANE}(c) shows \VANE measured in a sample featuring a \unit{50}{\nano\meter} thick Co$_2$FeAl film deposited on MgO substrate as a function of laser power $P$, with illumination on the central part of the main Hall bar in between the used contacts 2 and 4, and the magnetic field applied perpendicular to the main Hall bar. We observe the expected linear dependence of \VANE on $P$.

These calculations show that a scannable, focussed laser beam impinging on either ferromagnet or ferromagnet/normal metal thin film samples can be used for a controllable thermal landscape generation with substantial, laterally confined, temperature gradients along the sample normal.

\section{Magnetocaloritronics in other materials and geometries}
\begin{figure}
  \includegraphics[]{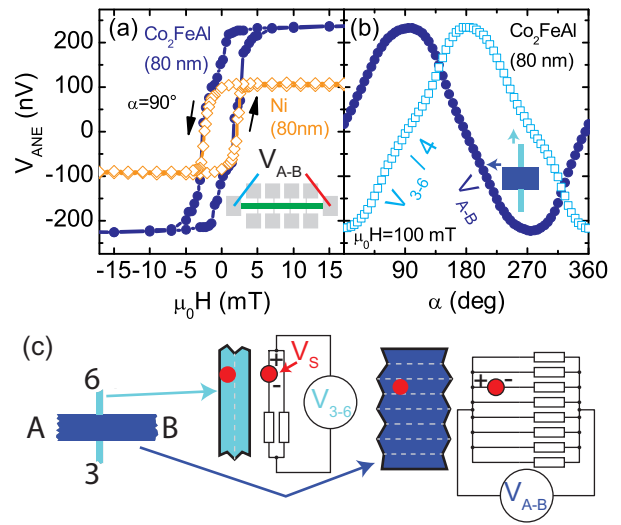}\\
  \caption{(a) $\VANE=V_\mathrm{A-B}$ averaged within the region indicated in the inset as a function of $\mu_0H$ in two further samples. (b) \VANE as a function of $\alpha$ for illumination on the $\unit{80}{\micro\meter}$ wide central Hall bar (solid symbols) and for illumination on the $\unit{20}{\micro\meter}$ wide contact line (open symbols). (c) Equivalent circuits of the main Hall bar (dark shading) and contact line (light shading) with parallel resistors and ideal voltage source $V_\mathrm{s}$ at the laser spot position (solid circle).}\label{fig:materials}
\end{figure}
To demonstrate the general nature of spin caloritronics, we carried out magnetocaloritronic measurements in a set of further samples. We observed \VANE in thin films of all three elemental ferromagnets, iron, cobalt and nickel and recorded \VISHE in a further GGG/YIG/Pt sample. Exemplarily, in Fig.~\ref{fig:materials}(a) we compare data obtained at an incident laser power $P=\unit{9}{\milli\watt}$ and a $\bm{H}$ orientation $\alpha=90\degree$ in a \unit{80}{\nano\meter} thick Ni film with data obtained in a \unit{80}{\nano\meter} thick Co$_2$FeAl sample, both patterned into the Hall bar geometry already introduced in Fig.~\ref{fig:spatial} in the main text.  The data correspond to \VANE averaged for illumination within a region of interest (ROI) on the main Hall bar as indicated by the inset. While the shapes of the \VANE hysteresis loops reflect the thin films' respective magnetic anisotropy, the magnitude of \VANE is approximately twice as large in the Co$_2$FeAl thin film. This is attributed mainly to the larger saturation magnetization $M_\mathrm{s}=\unit{1050}{\kilo\ampere\per\meter}$ of Co$_2$FeAl as opposed to $M_\mathrm{s}=\unit{370}{\kilo\ampere\per\meter}$ in Ni.

In the main text, Eqs.~(1) and (2) describe the generation of a \textit{local} magnetothermal electric field. This electric field is experimentally probed by an \textit{integral} voltage measurement which depends on sample geometry. Exemplarily we here show the dependence of \VANE on Hall bar width and orientation. To this end, we measured \VANE simultaneously at two contact pairs on the \unit{80}{\nano\meter} thick Co$_2$FeAl film sample as a function of $\alpha$. We used the orthogonal contact pairs A and B as well as 3 and 6 to obtain $V_\mathrm{A-B}$ and $V_{3-6}$, respectively. The results are presented in Fig.~\ref{fig:materials}(b) and correspond to the measurement of the projection of $\bm{E}$ to the $\bm{x}$ and $\bm{y}$ axis, respectively. The data are again averaged for illumination within the corresponding ROIs indicated in the inset in Fig.~\ref{fig:materials}(b). While $V_\mathrm{A-B} \propto \sin(\alpha)$ (closed symbols), $V_{3-6} \propto -\cos(\alpha)$ (open symbols). This angular dependency of \VANE again corroborates Eq.~\eqref{eq:ANE}. Turning to the \VANE magnitude in Fig.~\ref{fig:materials}(b), we find that the magnitude of $V_{3-6}(\alpha=180\degree)$ is exactly four times the magnitude of $V_\mathrm{A-B}(\alpha=90\degree)$. The Hall bar width $w$ in the region probed by $V_\mathrm{A-B}$ (dark shading) was \unit{80}{\micro\meter}, while that in the region probed by $V_{3-6}$ (light shading) was \unit{20}{\micro\meter}. The voltage $V$ observed in magnetocaloritronic experiments thus scales with $w^{-1}$. This can be understood in a simple model considering the magnetothermal generation of a local electromagnetic force at the laser spot position shunted by the unperturbed (non-illuminated) Hall bar cross-section as illustrated in Fig.~\ref{fig:materials}(c).  We assume the locally generated voltage $V_\mathrm{s}=E\cdot d$ to be homogeneous within the laser spot width $d=\unit{10}{\micro\meter}$ for simplicity. The Hall bar resistance is then modeled by a series of parallel resistances, at the position of one of which $V_\mathrm{s}$ is generated. The detected voltage $V$ thus is:
\begin{equation}\label{eq:width}
    V=\frac{d}{w} V_\mathrm{s}\;.
\end{equation}
This expression is, in particular, independent of the resistivity of the thin film (the resistance $R$ of all resistors sketched in Fig.~\ref{fig:materials}(c) and of Hall bar length. The magnitude of $V$ for a given thermal landscape is inversely proportional to the structure width $w$.  As $V$ increases with decreasing structure size, the technique described in this work is thus particularly applicable to micro- or nanoscale devices. Using Eq.~\eqref{eq:width}, we can now calculate the electric fields generated in our measurements. For the ANE measurements in Ni we obtain $E_\mathrm{ANE}=V_\mathrm{s}/d=\frac{w}{d^2}\cdot\VANE=\unit{87}{\milli\volt\per\meter}$ with $\VANE \approx \unit{109}{\nano\volt}$ taken from Fig.~\ref{fig:materials}(a). Accordingly, for Co$_2$FeAl we obtain $E_\mathrm{ANE}=\unit{176}{\milli\volt\per\meter}$ (cf. Fig.~\ref{fig:materials}(a)). For the \VISHE measurements, identical considerations apply and we obtain $E_\mathrm{ISHE}=\frac{w}{d^2}\cdot\VISHE=\unit{512}{\milli\volt\per\meter}$ (cf. Fig.~\ref{fig:integral}).

\begin{figure}
  \includegraphics[]{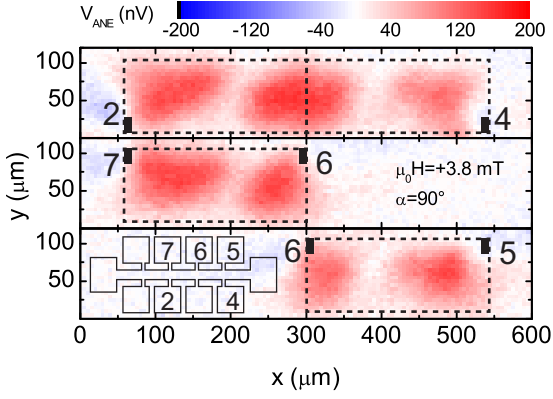}\\
  \caption{\VANE at \unit{+3.8}{\milli\tesla} in a sample with a \unit{50}{\nano\meter} thick Co$_2$FeAl film for three different sets of contacts. \VANE vanishes for illumination on either side of both contacts. Magnetic microstructure and \VANE magnitude is independent of the contact pair used.}\label{fig:spatial_s}
\end{figure}
The magnitude of the observed magnetothermal voltage is expected to be independent on Hall bar length or contact separation. This is demonstrated using the sample already investigated in Fig.~\ref{fig:spatial}. Here, Fig.~\ref{fig:spatial_s} shows data obtained via three simultaneous \VANE measurements using three different sets of contacts at constant magnetic field $\mu_0H=\unit{+3.8}{\milli\tesla}$. The top panel (contacts 2 and 4) is identical to that shown in Fig.~\ref{fig:spatial} for $\mu_0H=\unit{+3.8}{\milli\tesla}$. The second (contacts 7 and 6 - middle panel) and third (contacts 6 and 5 - bottom panel) set of contacts are both located on the side of the main Hall bar opposite to the first contact pair and are separated by \unit{230}{\micro\meter}. Clearly, \VANE is finite only if the laser beam and thus the vertical temperature gradient is positioned within the area enclosed by the used contacts. In any case, the observed magnetic microstructure and the magnitude of \VANE are independent of the selected contact pair. These observations provide clear evidence that the electric field is generated only locally at the laser spot position, since \VANE does \textit{not} depend on the contact separation.

\section{Magnetotransport in YIG/Pt}
\begin{figure}
  \includegraphics[]{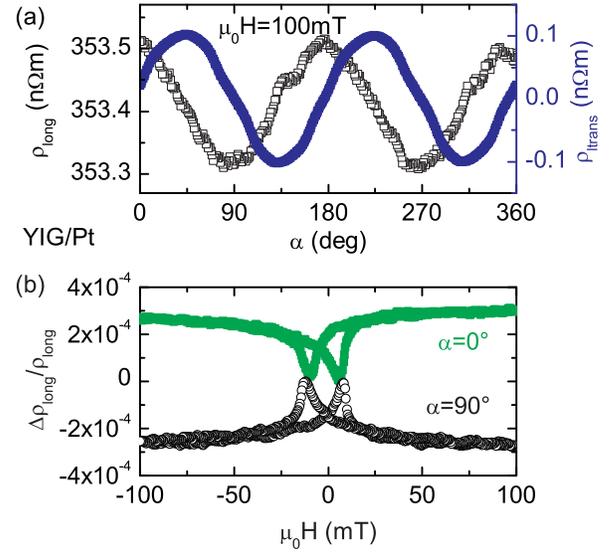}\\
  \caption{(a) ADMR measurement of $\rho_\mathrm{long}$ (left scale, open squares) and $\rho_\mathrm{trans}$ (right scale, solid circles) with $\mu_0H=\unit{100}{\milli\tesla}$ reveals an AMR of $\Delta\rho/\rho\approx 7\times 10^{-4}$. (b) $\rho_\mathrm{long}$ as a function of $H$ for $\alpha=90\degree$ (open circles) and $\alpha=0\degree$ (solid squares) shows coercive fields identical to YIG.}\label{fig:MR}
\end{figure}
We measured the resistivity of GGG/YIG/Pt samples as a function of the external magnetic field strength and orientation. Our pure YIG films (GGG/YIG(\unit{10}{\nano\meter})) are electrically insulating with a resistivity in excess of \unit{20}{\ohm\meter}. However, the electrical resistivity of the YIG/Pt sample is dependent on the external magnetic field. We determined $V_{1-4}$ (longitudinal voltage) and $V_{3-6}$ (transversal voltage) in a 4 point magnetotransport measurement with currents $I_\mathrm{A-B}$ ranging from \unit{0.1}{\milli\ampere} to \unit{2}{\milli\ampere}. The calculated resistivities $\rho_\mathrm{long}$ and $\rho_\mathrm{trans}$ obtained in a measurement with $I=\unit{1}{\milli\ampere}$ along contacts A-B are shown as a function of magnetic field orientation (ADMR measurement~\cite{Limmer:2006}) in Fig.~\ref{fig:MR}(a). Clearly, $\rho_\mathrm{long}\propto\cos^2(\alpha)$ (open symbols, left scale) and $\rho_\mathrm{trans}\propto\sin(2\alpha)$ (solid symbols, right scale) indicative of anisotropic magnetoresistance (AMR)~\cite{McGuire:1975}. Furthermore, as shown in Fig.~\ref{fig:MR}(b), $\rho_\mathrm{long}$ shows switching fields coinciding with the coercive field $\mu_0H_\mathrm{c}\approx\unit{10}{\milli\tesla}$ of our YIG thin film. In accordance to the expectations for AMR, the switching changes direction under a rotation of $\mathbf{H}$ of $90\degree$. This can be observed in Fig.~\ref{fig:MR}(b) where the curve recorded for $\alpha=0\degree$ appears reflected along the horizontal axis with respect to the curve obtained at $\alpha=90\degree$. The total AMR amounts to $\Delta\rho/\rho_\mathrm{long}\approx 7\times 10^{-4}$, about a factor of 10 lower than usually observed in elemental conductive ferromagnets~\cite{McGuire:1975}.  The AMR ratio was independent of $I$ in the range of $\unit{0.1}{\milli\ampere}\leq I \leq \unit{2}{\milli\ampere}$, ruling out thermoelectric effects due to Joule heating as the cause of the magnetoresistance. We attribute the magnetoresistance to induced magnetic moments in the Pt layer close to the YIG/Pt interface. This proximity effect is expected for clean interfaces within the first few monolayers ($\approx \unit{1}{\nano\meter}$) of Pt~\cite{Wilhelm:2000, Meier:2011}, explaining the small AMR. Because of the identical symmetry of the longitudinal SSE and the ANE, one needs to consider a possible contribution of the magnetized Pt to an ANE. Assuming that the entire Pt layer is magnetized with a magnetization identical to YIG, ($M_\mathrm{s}^\mathrm{YIG}=\unit{127}{\kilo\ampere\per\meter}$ obtained by SQUID magnetometry), and using the calculated mean temperature gradient in the Pt film, $\nabla T=\unit{-0.6}{\kelvin\per\micro\meter}$ (cf. Fig.~\ref{fig:gradTSSE}), an anomalous Nernst coefficient $N_\mathrm{Pt}=\unit{5.3\times10^{-6}}{\volt\per\kelvin\tesla}$ needs to be assumed to explain $\VISHE=\unit{640}{\nano\volt}$ (cf. Fig.~\ref{fig:integral}) based on an ANE in magnetized Pt alone. This coefficient is more than one order of magnitude larger than the ANE coefficient of Ni and Co$_2$FeAl even with the unrealistic assumption of an entirely magnetized Pt film. Furthermore this Nernst coefficient would be two orders of magnitude larger than the Nernst coefficient of Pt, $N_\mathrm{Pt}=\unit{1.3\times10^{-8}}{\volt\per\kelvin\tesla}$~\cite{Lueck:1964}, ruling out a Nernst effect in the Pt layer in the stray field of the YIG thin film as the origin of \VISHE. Assuming $N_\mathrm{Pt}=N_\mathrm{Ni}=\unit{1.3\times10^{-7}}{\volt\per\kelvin\tesla}$ and that only half of the Pt film is magnetized to $M_\mathrm{s}^\mathrm{YIG}$ and with $\nabla T=\unit{-0.7}{\kelvin\per\micro\meter}$ averaged in the first \unit{3.5}{\nano\meter} of Pt at the YIG interface, we can calculate the expected maximum voltage due to an ANE as $\VANE\approx\unit{9}{\nano\volt}$ for the YIG/Pt measurement, as opposed to $\VISHE\approx\unit{640}{\nano\volt}$ measured. Hence, we conclude that Nernst effects due to the proximity effect in Pt are at most a small contribution ($<2\%$) to \VISHE and we thus attribute \VISHE to the longitudinal spin Seebeck effect. This is furthermore supported by the calculated spin Seebeck coefficient $S=\unit{5.9\times10^{-8}}{\volt\per\kelvin}$ which is in accordance to literature. Further studies of proximity effects in Pt/ferromagnet samples are however mandatory for a full quantitative understanding of spin Seebeck effects detected via the inverse spin Hall effect, as even in the conventional spin Seebeck geometry spurious temperature gradients exist and are known to give rise to anomalous Nernst effects~\cite{Bosu:2011}.


\begin{thebibliography}{10}

\bibitem{Nernst:1887}
W.~Nernst, Annalen der Physik \textbf{267}, 760 (1887).

\bibitem{Harmann:1967}
T.~Harmann and J.~M. Honig, \emph{Thermoelectric and Thermomagnetic Effects and
  Applications} (McGraw-Hill, New York, 1967).

\bibitem{Vasil'eva:1972}
R.~P. Vasil'eva and B.~Akmuradov, Russ. Phys. J. \textbf{15}, 814 (1972).

\bibitem{Miyasato:2007}
T.~Miyasato, N.~Abe, T.~Fujii, A.~Asamitsu, S.~Onoda, Y.~Onose, N.~Nagaosa, and
  Y.~Tokura, Phys. Rev. Lett. \textbf{99}, 086602 (2007).

\bibitem{Suryanarayanan:1999}
R.~Suryanarayanan, V.~Gasumyants, and N.~Ageev, Phys. Rev. B \textbf{59}, R9019
  (1999).

\bibitem{Lee:2004}
W.-L. Lee, S.~Watauchi, V.~L. Miller, R.~J. Cava, and N.~P. Ong, Phys. Rev.
  Lett. \textbf{93}, 226601 (2004).

\bibitem{Hanasaki:2008}
N.~Hanasaki, K.~Sano, Y.~Onose, T.~Ohtsuka, S.~Iguchi, I.~K\'ezsm\'arki,
  S.~Miyasaka, S.~Onoda, N.~Nagaosa, and Y.~Tokura, Phys. Rev. Lett.
  \textbf{100}, 106601 (2008).

\bibitem{Pu:2008}
Y.~Pu, D.~Chiba, F.~Matsukura, H.~Ohno, and J.~Shi, Phys. Rev. Lett.
  \textbf{101}, 117208 (2008).

\bibitem{Uchida:2008}
K.~Uchida, S.~Takahashi, K.~Harii, J.~Ieda, W.~Koshibae, K.~Ando, S.~Maekawa,
  and E.~Saitoh, Nature \textbf{455}, 778 (2008).

\bibitem{Hirsch:1999}
J.~E. Hirsch, Phys. Rev. Lett. \textbf{83}, 1834 (1999).

\bibitem{Saitoh:2006}
E.~Saitoh, M.~Ueda, H.~Miyajima, and G.~Tatara, Appl. Phys. Lett. \textbf{88},
  182509 (2006).

\bibitem{Uchida2:2010}
K.~Uchida, H.~Adachi, T.~Ota, H.~Nakayama, S.~Maekawa, and E.~Saitoh, Appl.
  Phys. Lett. \textbf{97}, 172505 (2010).

\bibitem{Bosu:2011}
S.~Bosu, Y.~Sakuraba, K.~Uchida, K.~Saito, T.~Ota, E.~Saitoh, and K.~Takanashi,
  Phys. Rev. B \textbf{83}, 224401 (2011).

\bibitem{Jaworski:2010}
C.~M. Jaworski, J.~Yang, S.~Mack, D.~D. Awschalom, J.~P. Heremans, and R.~C.
  Myers, Nat. Mater. \textbf{9}, 898 (2010).

\bibitem{Uchida:2010}
K.~Uchida, J.~Xiao, H.~Adachi, J.~Ohe, S.~Takahashi, J.~Ieda, T.~Ota,
  Y.~Kajiwara, H.~Umezawa, H.~Kawai, G.~E.~W. Bauer, S.~Maekawa, and E.~Saitoh,
  Nat. Mater. \textbf{9}, 894 (2010).

\bibitem{Gravier:2005}
L.~Gravier, S.~Serrano-Guisan, and J.-P. Ansermet, J. Appl. Phys. \textbf{97},
  10C501 (2005).

\bibitem{Gravier:2006}
L.~Gravier, S.~Serrano-Guisan, F.~Reuse, and J.-P. Ansermet, Phys. Rev. B
  \textbf{73}, 024419 (2006).

\bibitem{Slonczweski:2010}
J.~C. Slonczewski, Phys. Rev. B \textbf{82}, 054403 (2010).

\bibitem{Yu:2010}
H.~Yu, S.~Granville, D.~P. Yu, and J.-P. Ansermet, Phys. Rev. Lett.
  \textbf{104}, 146601 (2010).

\bibitem{Slachter:2010}
A.~Slachter, F.~L. Bakker, J.~Adam, and B.~J. van Wees, Nat. Phys. \textbf{6},
  879 (2010).

\bibitem{leBreton:2011}
J.-C. Le~Breton, S.~Sharma, H.~Saito, S.~Yuasa, and R.~Jansen, Nature
  \textbf{475}, 82 (2011).

\bibitem{Reichling:1994}
M.~Reichling and H.~Gr\"{o}nbeck, J. Appl. Phys. \textbf{75}, 1914 (1994).

\bibitem{Weiler:2011}
M.~Weiler, F.~D. Czeschka, A.~Brandlmaier, I.-M. Imort, G.~Reiss, A.~Thomas,
  G.~Woltersdorf, R.~Gross, and S.~T.~B. Goennenwein, Appl. Phys. Lett.
  \textbf{98}, 042501 (2011).

\bibitem{Ri:1994}
H.-C. Ri, R.~Gross, F.~Gollnik, A.~Beck, R.~P. Huebener, P.~Wagner, and
  H.~Adrian, Phys. Rev. B \textbf{50}, 3312 (1994).

\bibitem{Wilhelm:2000}
F.~Wilhelm, P.~Poulopoulos, G.~Ceballos, H.~Wende, K.~Baberschke,
  P.~Srivastava, D.~Benea, H.~Ebert, M.~Angelakeris, N.~K. Flevaris,
  D.~Niarchos, A.~Rogalev, and N.~B. Brookes, Phys. Rev. Lett. \textbf{85}, 413
  (2000).

\bibitem{Limmer:2006}
W.~Limmer, M.~Glunk, J.~Daeubler, T.~Hummel, W.~Schoch, R.~Sauer, C.~Bihler,
  H.~Huebl, M.~S. Brandt, and S.~T.~B. Goennenwein, Phys. Rev. B \textbf{74},
  205205 (2006).

\bibitem{Weiler:2009}
M.~Weiler, A.~Brandlmaier, S.~Gepr\"{a}gs, M.~Althammer, M.~Opel, C.~Bihler,
  H.~Huebl, M.~S. Brandt, R.~Gross, and S.~T.~B. Goennenwein, New J. Phys.
  \textbf{11}, 013021 (2009).

\bibitem{Smith:1911}
A.~W. Smith, Phys. Rev. (Series I) \textbf{33}, 295 (1911).

\bibitem{Avery:2011}
A.~D. Avery, R.~Sultan, D.~Bassett, D.~Wei, and B.~L. Zink, Phys. Rev. B
  \textbf{83}, 100401 (2011).

\bibitem{Slack:1971}
G.~A. Slack and D.~W. Oliver, Phys. Rev. B \textbf{4}, 592 (1971).

\bibitem{Palik:1985}
E.~D. Palik, editor, \emph{Handbook of Optical Constants of Solids} (Academic
  Press, San Diego, 1985), 5th ed.

\bibitem{Vargaftik:1993}
N.~B. Vargaftik, \emph{Handbook of Thermal Conductivity of Liquids and Gases}
  (CRC Press, Boca Raton, 1993), 1st ed.

\bibitem{Farrell:1969}
T.~Farrell and D.~Greig, J. Phys. C: Solid State Phys. \textbf{2}, 1465 (1969).

\bibitem{Itatani:2006}
K.~Itatani, T.~Tsujimoto, and A.~Kishimoto, J. Eur. Ceram. Soc. \textbf{26},
  639  (2006).

\bibitem{Slocum:1956}
E.~W. Slocum, editor, \emph{Tables of thermal properties of gases} (National
  Bureau of Standards, Washington, 1956), 1st ed.

\bibitem{Lide:2003}
D.~R. Lide, editor, \emph{CRC Handbook of Chemistry and Physics} (CRC Press,
  Boca Raton, 2003), 84th ed.

\bibitem{Watanabe:1993}
H.~Watanabe, Thermochim. Acta \textbf{218}, 365  (1993).

\bibitem{McGuire:1975}
T.~R. McGuire and R.~I. Potter, IEEE Trans. Magn. \textbf{11}, 1018 (1975).

\bibitem{Meier:2011}
F.~Meier, S.~Lounis, J.~Wiebe, L.~Zhou, S.~Heers, P.~Mavropoulos, P.~H.
  Dederichs, S.~Bl\"ugel, and R.~Wiesendanger, Phys. Rev. B \textbf{83}, 075407
  (2011).

\bibitem{Lueck:1964}
R.~L\"{u}ck and T.~Ricker, phys. status solidi (b) \textbf{7}, 817 (1964).

\end{thebibliography}
\end{document}